\documentclass[twocolumn,amsmath,amssymb,prl,10pt,nofootinbib,superscriptaddress]{revtex4}
\usepackage{ graphicx, float,amsmath, amsmath, amssymb}
\usepackage[export]{adjustbox}

\def\be{\begin{equation}}
\def\ee{\end{equation}}
\def\bea{\begin{eqnarray}}
\def\eea{\end{eqnarray}}
\def\bse{\begin{subequations}}
\def\ese{\end{subequations}}

\usepackage[breaklinks, colorlinks, citecolor=blue]{hyperref}
\usepackage[labelsep=period]{caption}
\begin{document}
\title{Fast radio bursts and the stochastic lifetime of black holes in quantum gravity}
\author{Aur\'elien Barrau}%
\affiliation{%
Laboratoire de Physique Subatomique et de Cosmologie, Universit\'e Grenoble-Alpes, CNRS/IN2P3\\
53, avenue des Martyrs, 38026 Grenoble cedex, France
}
\author{Flora Moulin}%
\affiliation{%
Laboratoire de Physique Subatomique et de Cosmologie, Universit\'e Grenoble-Alpes, CNRS/IN2P3\\
53, avenue des Martyrs, 38026 Grenoble cedex, France
}

\author{Killian Martineau}%
\affiliation{%
Laboratoire de Physique Subatomique et de Cosmologie, Universit\'e Grenoble-Alpes, CNRS/IN2P3\\
53, avenue des Martyrs, 38026 Grenoble cedex, France
}
\date{\today}
\begin{abstract} 
Nonperturbative quantum gravity effects might allow a black-to-white hole transition. We revisit this increasingly popular hypothesis by taking into account the fundamentally random nature of the bouncing time. We show that if the primordial mass spectrum of black holes is highly peaked, the expected signal can in fact match the wavelength of the observed fast radio bursts. On the other hand, if the primordial mass spectrum is wide and smooth, clear predictions are suggested and the sensitivity to the shape of the spectrum is studied.
 \end{abstract}
\maketitle

\section{Introduction}

Finding observational consequences of quantum gravity is obviously a major challenge. In the last decade most attempts have focused on the early Universe, evaporating black holes or Lorentz invariance violation (see \cite{Barrau:2017tcd} for a recent overview). In the last years, the idea that quantum gravity effects could be seen in higher-mass black holes has attracted a lot of interest \cite{Rovelli:2014cta,Haggard:2014rza,Barcelo:2015uff,Barcelo:2017lnx,Haggard:2016ibp}. In particular it was suggested that the quite mysterious Fast Radio Bursts (FRBs)  \cite{Katz:2016dti} could be explained by bouncing black holes \cite{Barrau:2014yka}. There are unquestionably simpler astrophysical explanations that we consider to be more probable but this hypothesis is worth a deeper look. At the heuristic and intuitive level, this bounce can be understood as a phenomenon quite similar to what is expected to happen to the Universe in loop quantum cosmology \cite{lqc9,Barrau:2017rwl}. In the cosmological framework, the classically contracting branch is linked to the classically expanding one by a quantum tunneling, whereas in the black hole sector, the classically collapsing solution is glued to the classically exploding one (on the double cover of the Kruskal map \cite{Haggard:2014rza}). The usual event horizon is replaced by a trapping horizon \cite{Ashtekar:2004cn}. In this brief article we revisit this hypothesis by taking into account the fundamental randomness of the tunneling process that was previously ignored. In the first section we assume a peaked mass spectrum for the bouncing black holes and show that the 3 orders of magnitude in energy thought to be missing to explain FRBs can easily be accounted for. In the second section we consider a wide mass spectrum and investigate the sensitivity of the signal to the spectral index. We show that the expected emission remains compatible with measurements and make clear predictions. 

\section{Peaked mass spectrum}

The heuristic arguments given by Rovelli, Haggard and Vidotto in the previously mentioned articles suggested that the black hole lifetime could be of the order of $M^2$ in Planck units (those units are used throughout the rest of the article except otherwise stated). As this is shorter that the Hawking evaporation time (of the order of $M^3$) this means that black holes might bounce before they evaporate: the Hawking effect would just be a dissipative correction. An exact calculation of this lifetime is in principle possible in loop quantum gravity (see, {\it e.g.}, \cite{Rovelli:2011eq}), but it is still hard to perform accurately at this stage \cite{Christodoulou:2016vny}. The previous phenomenological works around this hypothesis have focused on gamma-ray bursts \cite{Barrau:2014hda}, FRBs \cite{Barrau:2014yka}, the space-integrated signal \cite{Barrau:2015uca}, and trying to explain the Fermi excess \cite{Barrau:2016fcg}. In all of them the lifetime was taken, as a first approximation, to be deterministic, fixed at the value $\tau=kM^2$ where $k$ was chosen to be of the order of 0.05 (however in one of the studies \cite{Barrau:2015uca} its value was varied). We also assume this value in the present article as it the most phenomenologically interesting one (and the smallest one theoretically allowed). However, as the black-to-white hole transformation is to be understood as a tunneling process, the lifetime of a black hole should be considered as a random variable.\\

The probability that a black hole has not yet bounced after a time $t$ is given by
\begin{equation}
P(t)=\frac{1}{\tau}e^{-\frac{t}{\tau}}.
\end{equation}
This is the usual ``nuclear decay" behavior which  comes directly from the fact that the number of bouncing black holes during a time interval $dt$ is proportional to the full number of black holes and to $dt$. 
We focus in this study on local effects and neglect the redshift integration as this will play only a minor role in the analysis carried out. The black holes we are interested in can be considered to have been produced in the early universe as the range of masses -- much below a Solar mass - leading to bounces occurring in the contemporary universe can only be associated with primordial black holes (PBHs, see \cite{Carr:2009jm} for a quite recent review on the limits on the PBH abundance and references therein for possible formation mechanisms. In general, the number of black holes of a given type bouncing after a time $t_H$ (taken to be the Hubble time as we are considering phenomena taking place nowadays) in a time interval $dt$ is:
\begin{equation}
dN=\frac{N_0}{kM^2}e^{-\frac{t_H}{kM^2}}dt,
\end{equation}

where $N_0$ is the initial abundance. The exponential function entering this calculation directly comes the random nature of the bounce, as in the previous formula.
Let us assume that the initial differential mass spectrum of the considered PBHs is given by $dN/dM$.\\

In this study, we focus on the so-called bouncing black hole {\it low-energy} component as this is the one that is relevant for a possible link with FRBs. This specific component is based on a simple dimensional analysis : photons are assumed to be emitted with a characteristic wavelength that is of the order of the size of the black hole, which is the only length scale of the problem. As in \cite{Barrau:2016fcg}, we model the shape of the signal emitted by a single black hole by a simple Gaussian function:
\begin{equation}
\frac{dN_{\gamma}^{BH}}{dE}=Ae^{-\frac{(E-E_0)^2}{2\sigma_E^2}},
\end{equation}
where $E_0=1/(2R_S)=1/(4M)$, $R_S$ is the Schwarzschild radius and $M$  is the mass of the considered black hole. This choice is arbitrary and simply taken as an example. The width is typically fixed to be $\sigma_E=0.1E_0$ but the results do not critically depend  on this value or the detailed shape of the distribution.\\

The full signal due to a local distribution of bouncing black holes is given by 
\begin{equation}
\frac{dN_{\gamma}}{dE}=\int_{M_{Pl}}^{\infty}Ae^{-\frac{(E-E_0)^2}{2\sigma_E^2}}\cdot \frac{dN}{dM}(M)\cdot\frac{1}{kM^2}e^{-\frac{t_H}{kM^2}}.
\end{equation}

The point we want to raise in this study is that the mean energy of the detected signal might {\it not} be the naively expected one, that is may {\it not} be $E\sim1/(4M_{t_H})$ where $M_{t_H}$ is the mass satisfying $t_H=kM_{t_H}^2$ (this would correspond to black holes having a characteristic lifetime equal to the age of the Universe). The naive expectation, $E\sim1/(4M_{t_H})$ is not in the radio band, but rather 3 orders of magnitude higher in energy, in the infrared band. If the initial mass spectrum is peaked around a value $M_0$, {\it e.g.}, according to 
\begin{equation}
\frac{dN}{dM}\propto e^{-\frac{(M-M_0)^2}{2\sigma_{\footnotesize{M}}^2}},
\end{equation}
which can in principle be different than $\sqrt{t_H/k}$, the energy will however be peaked around $1/(4M_0)$ which can differ from $1/(4M_{t_H})$. This is  possible precisely because of the distributional nature of the actual bouncing time.\\

Considering a peaked mass spectrum is not arbitrary and can be justified if PBHs are created, for example, because of a phase transition in the early Universe (see, {\it e.g.}, \cite{Sobrinho:2016fay}). As the primordial cosmological power spectrum is now clearly known not to be blue \cite{Ade:2015xua} (at least at large scales), the naturally expected density contrast is not high enough to produce PBHs \cite{Carr:1975qj} and specific post-inflationary phenomena are generically required (see, {\it e.g.}, \cite{Jedamzik:1999am}). 


\begin{figure}[H]
\includegraphics[width=85mm,center]{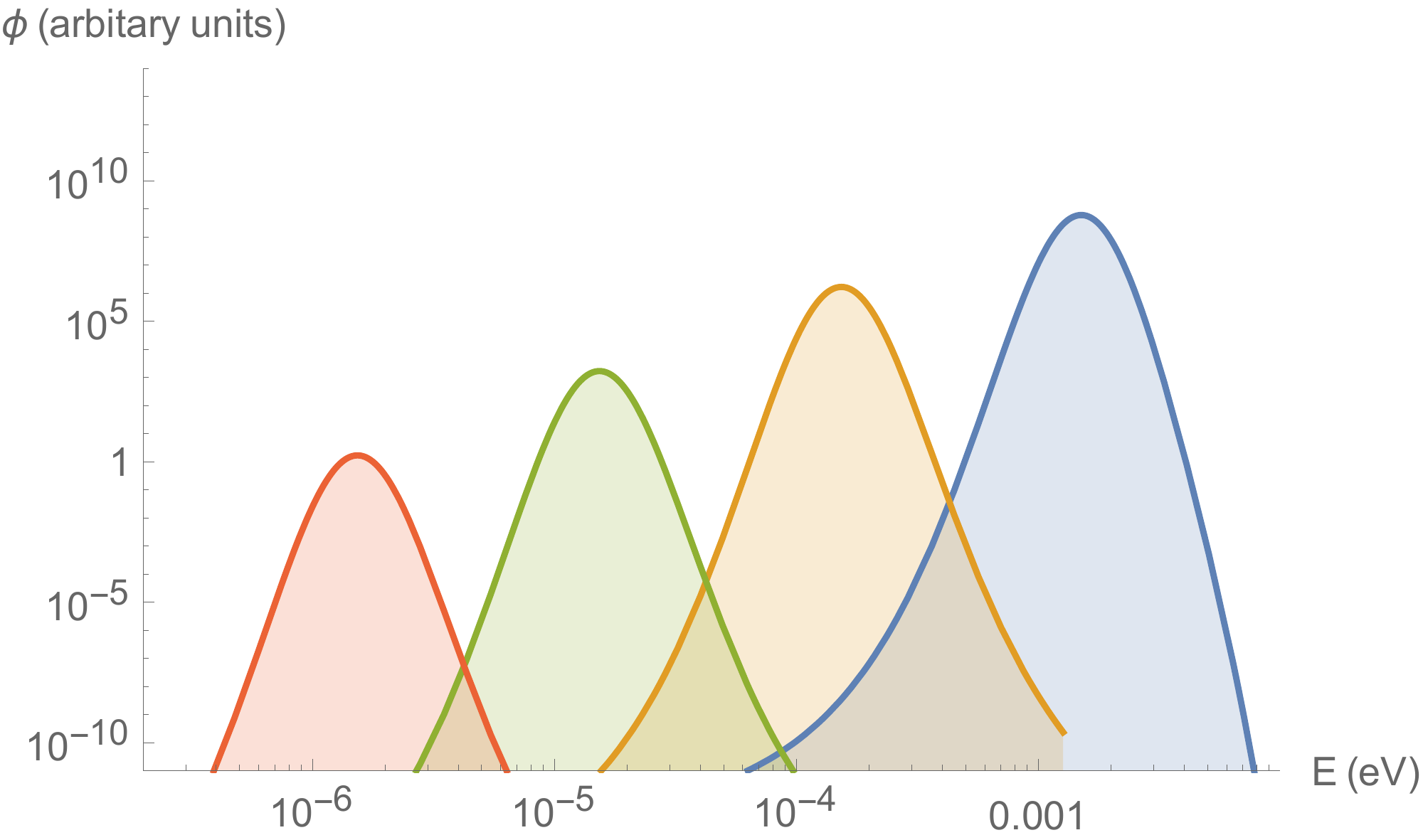}
\caption{Differential electromagnetic flux emitted by bouncing PBHs for a central mass $M_0$ equal (from right to left) to $M_{t_H}$, $10M_{t_H}$, $100M_{t_H}$, and $1000M_{t_H}$. The normalization is such that the total mass going into PBHs is the same in all cases.}
\label{f2}
\end{figure}

In Fig \ref{f2}, the expected emitted flux is shown for different values of the central mass $M_0$ of the initial mass spectrum: $M_{t_H}$, $10M_{t_H}$, $100M_{t_H}$, and $1000M_{t_H}$. As expected, this shows that the energy of the signal depends on the mass spectrum even if the parameters of the model are fixed. Naturally, when the mass spectrum is peaked at masses well above $M_{t_H}$, the amplitude of the expected signal decreases as BHs that are exploding today constitute an increasingly smaller fraction of the full population. However, the key point we stress here is that a given mean lifetime $\tau=kM^2$ does not imply a fixed expected energy.\\

In particular, it was previously emphasized that the expected mean wavelength (obtained by fixing $\tau=t_H$) of the electromagnetic emission associated with bouncing black holes was basically one thousand times smaller than required to explain the FRBs. If the mass spectrum is peaked at masses higher than $M_{t_H}$, it is however perfectly possible to precisely account for the expected wavelength. The curve on the left in Fig \ref{f2} is peaked around 1.5 GHz, which corresponds to the  typical wavelength of FRBs. At this stage, there is no obvious motivation for choosing a specific value for the peak mass. Interesting proposals were recently suggested, for example in the framework of critical Higgs inflation \cite{Ezquiaga:2017fvi}, but as pointed out in the mentioned reference, the actual peak value could differ from the naively calculated one by several orders of magnitude due to accretion and merging and many other models do exist that suggest other mass values. 

In Fig \ref{f2} the normalization between the different curves is such that the total mass going into black holes is the same:
\begin{equation}
\int_{M_{Pl}}^{\infty}M\frac{dN}{dM}=cte.
\end{equation}
This is somehow  justified if ones tries to account for dark matter with PBHs. The point we want to stress with this remark is simply  that the  decrease in flux when one moves below the ``natural" mass $M_{T_H}$ is not drastic.  Accounting for the observed events by shifting the peaked mass to higher values requires a higher density of PBHs. This cannot be done up to arbitrary values as the upper bounds on the density of PBHs would then be violated. However, orders of magnitude show that the density of PBHs required to account for observed events is very far below the known bounds and this does not limit the present proposal as the rate of FRBs is actually very small \cite{Fialkov:2017qoz}. There is no point is performing a detailed normalization of the expected spectrum at this stage as the initial mass spectrum normalization is totally unknown and the calculation of any observable would directly depend on it.

We have also considered a second normalization, such that the total number of black holes is the same,
\begin{equation}
\int_{M_{Pl}}^{\infty}\frac{dN}{dM}=cte,
\end{equation}
and this basically leads to the very same results. 
\\

Beyond FRBs -- which can be explained by astrophysical phenomena -- the point raised here is simply the fact that when the probabilistic nature of the bouncing time is accounted for, the mean energy of the emitted signal is also determined by the mass spectrum and not only by the lifetime of the black holes.

\section{Wide mass spectrum}

\begin{figure}[H]
\includegraphics[width=85mm,center]{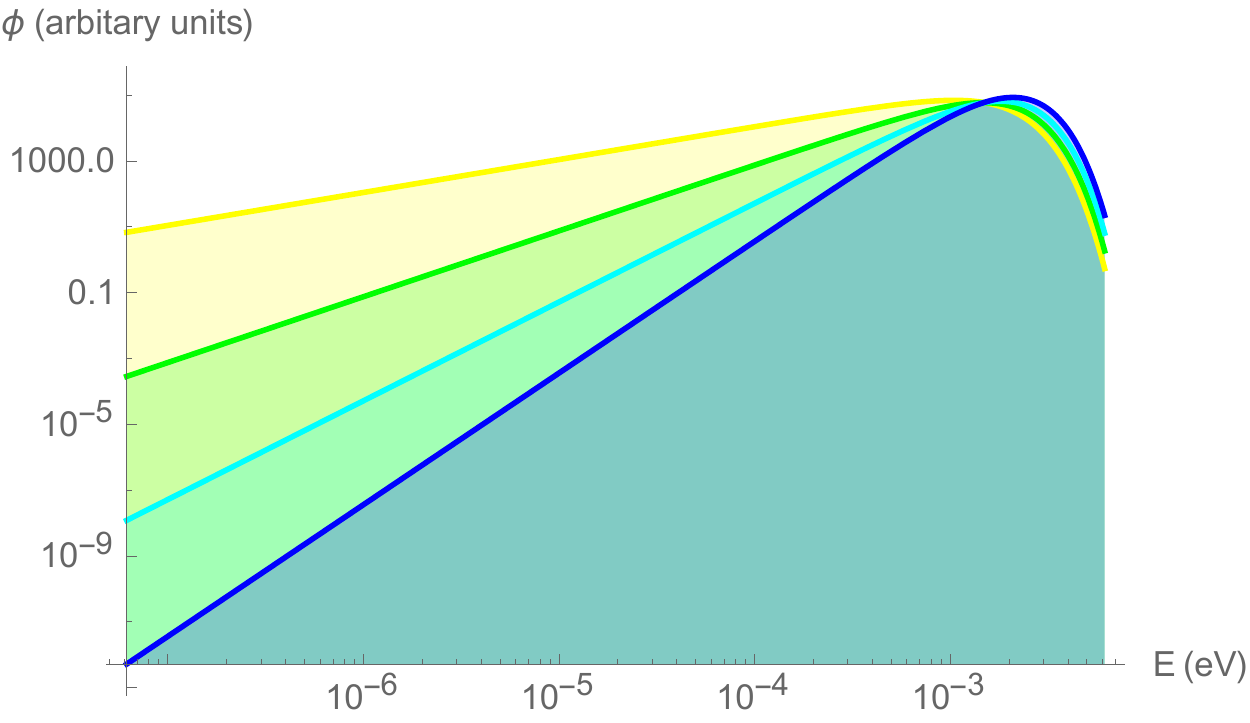}
\caption{Signal expected from a wide mass spectrum, with $\alpha=\{-3,-2,-1,0\}$ from the lower curve to the upper curve at $10^{-6}$ eV.}
\label{f3}
\end{figure}

It is also possible that the mass spectrum of PBHs is quite wide. As a toy model, if it is directly produced by scale-invariant density perturbations in a perfect fluid with equation of state $w=p/\rho$, the mass spectrum can be approximated by \cite{Carr:1975qj}
\begin{equation}
\frac{dN}{dM} \propto M^{-1-\frac{1+3w}{1+w}}.
\label{spec}
\end{equation}
In this study, we just consider -- as a first approximation -- a spectrum
\begin{equation}
\frac{dN}{dM} \propto M^{\alpha},
\label{spec}
\end{equation}
where $\alpha$ is an unknown parameter. In Fig \ref{f3} we present the expected signal for $\alpha=\{-3,-2,-1,0\}$ (a spectrum rising with an increasing mass on a wide interval would be rather unphysical). Once again, the shape of the mass spectrum does influence the expected signal as the probabilistic nature of the lifetime is now taken into account: black holes with masses smaller or larger than $M_{t_H}$ do also contribute to the emitted radiation and changing their relative weights does change the result.

This leads to another way of addressing the discrepancy between the ``natural" wavelength (around 0.02 cm $\sim2\times10^{-6}$ eV) of bouncing black holes and the observed wavelength (around 20 cm $\sim2\times10^{-3}$ eV) of FRBs. It could indeed be that most bouncing black holes do lead to a signal of wavelength $\sim$ 0.02 cm and that only the tail (which exists because of the probabilistic nature of the lifetime) of the distribution is observed in the radio band. If the peak is in the infrared -- which should occur if the mass spectrum is wide -- it might be that it is simply unobserved today. Detectors in the infrared band have proper time constants that are way to high to allow for the measurement of such fast transient phenomena and there are no deep surveys being carried out.\\

In this case, as shown in Fig \ref{f3}, a clear prediction of this model for future observations is that one should expect a higher flux as the energy increases (up to the infrared band). The slope of this increase reflects that of the mass spectrum. This is qualitatively quite independent of the details of the mass spectrum. 

\section{Conclusion}

The possible existence of a black-to-white hole transition through a kind of tunneling process has recently received a lot of attention in quantum gravity. In this brief article we have taken into account the fundamentally random nature of the black hole lifetime in those models. We showed that this can induce a substantial shift with respect to previous studies in which the characteristic lifetime $\tau$ (either derived from the full theory -- first attempts can be found in \cite{Christodoulou:2016vny} -- or inferred by heuristic arguments) was taken as an actual bouncing time. 

In a Poisson process, the distribution of time intervals is wide and exponentially decreasing. A bounce can occur after a time which is very different from its characteristic timescale, with the smallest time being always the most probable one. This should be taken into account (and this was indeed accounted for in \cite{Raccanelli:2017xee}).\\

Beyond this quite trivial statement, we have shown that, because of this stochastic process, the mean energy of the emitted signal can be different that previously considered. In particular, if the mass spectrum of PBHs is peaked, it is perfectly possible to match the observed FRBs.

In addition if the mass spectrum of PBHs is wide and continuous it is still possible to explain the data and a prediction was suggested for future observations.\\

The main point of this study was not to revive at any price the hypothesis that FRBs are due to bouncing black holes. Our point was to show that the randomness of the lifetime of black holes in quantum gravity can drastically change the spectral characteristic of the expected signal when the mass spectrum is highly peaked and can lead to interesting predictions in any case. 

\section{Acknowledgments}

K.M is supported by a grant from the CFM foundation.

\bibliography{refs}
 \end{document}